\documentclass[aps,prl,preprint,groupedaddress,floatfix]{revtex4-1}
\usepackage{bm}
\usepackage{graphicx}

\newcommand{\CSI}{CsSnI$_{\mbox{\scriptsize 3}}$}
\newcommand{\Ba}{B-$\alpha$}
\newcommand{\Bb}{B-$\beta$}
\newcommand{\Bg}{B-$\gamma$}
\def\<{\langle}
\def\>{\rangle}

\def\w{\omega}
\def\wall{\bm{\omega}}

\def\V{{\tilde{V}}}
\def\F{{\tilde{F}}}
\def\Ff{{\mathcal{F}}}
\def\SCw{$SC\omega$}

\let\oldhat\hat
\renewcommand{\vec}[1]{\mathbf{#1}}
\renewcommand{\hat}[1]{\oldhat{\mathbf{#1}}}

\begin{document}

\title{Anharmonic stabilization and band gap renormalization in the perovskite \CSI}

\author{Christopher E. Patrick}
\author{Karsten W. Jacobsen}
\author{Kristian S. Thygesen}
\affiliation{Center for Atomic-Scale Materials Design (CAMD), Department of Physics,
Technical University of Denmark, DK---2800 Kongens Lyngby, Denmark}


\begin{abstract}
Amongst the X(Sn,Pb)Y$_3$ perovskites currently under scrutiny for their 
photovoltaic applications,
the cubic \Ba \ phase of \CSI \ is  arguably the best characterized experimentally.
Yet, according to the standard harmonic theory of phonons,
this deceptively simple phase should not exist at all due to 
rotational instabilities of the SnI$_6$ octahedra.
Here, employing self-consistent phonon theory 
we show that these soft modes are stabilized at experimental conditions
through anharmonic phonon-phonon interactions between the Cs
ions and their iodine cages.
We further calculate the renormalization
of the electronic energies due to vibrations and find an unusual opening
of the band gap, estimated as 0.24 and 0.11~eV at 500 and 300~K, 
which we attribute to the stretching of Sn--I bonds.
Our work demonstrates the important role of temperature in accurately describing these materials.
\end{abstract}

\date{\today}

\maketitle

Four decades after its identification as an unusual phase-change 
material~\cite{Donaldson1973}, the inorganic perovskite \CSI \ has 
experienced a revival of interest in its technological applications.
After being used as a hole transporter in solid-state 
photovoltaics~\cite{Chung20122}, the subsequent explosion in activity 
surrounding perovskite 
solar cells~\cite{Lee2012} has seen \CSI \ incorporated into new devices
as a lead-free light absorber~\cite{Kumar2014} with favorable optical 
properties~\cite{Stoumpos2013,Chen2012,Shum2010,Yu2011}.
Like many perovskites~\cite{KingSmith1994} \CSI \ has
a rich phase diagram,
driven by low-energy rotations and tilts of the
SnI$_6$ octahedra~\cite{Yamada1991,Huang2014,Chung2012}.
In addition \CSI \ has an unusual electronic structure, with a non-degenerate
and highly-dispersive valence band~\cite{Huang2013}
and an intra-atomic band gap
strongly coupled to external strain~\cite{Huang2013,Li2015,Borriello2008}.

\begin{figure}
\includegraphics{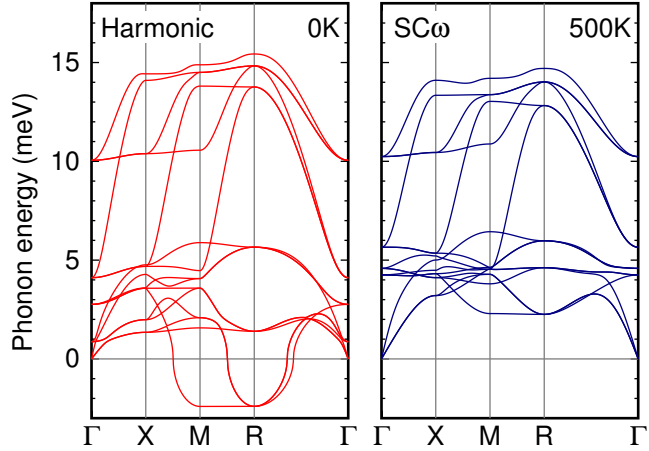}
\caption{
(color online)
Phonon bandstructures obtained for the \Ba \ phase of \CSI \ calculated under
the harmonic approximation or with self-consistent phonon frequencies (\SCw).
Imaginary frequencies are shown as negative.
The supercell calculations do not include the non-analytic correction
accounting for the long-wavelength splitting of polar modes~\cite{Wang2012}.
\label{fig.harm_comp}
}
\end{figure}

In a wider context,
\CSI \ is the gateway to understanding the
basic physics of the family of X(Sn,Pb)Y$_3$ perovskites (X = cation, Y = halogen).
Unlike its famous cousin MAPbI$_3$ (MA = methylammonium),
\CSI \ has 
(i) no permanent cationic dipole moment~\cite{Frost2014},
(ii) reduced spin-orbit coupling due to the lighter mass of Sn~\cite{Umari2014}
and 
(iii) a high-symmetry cubic (\Ba) phase characterized
by many studies~\cite{Scaife1974,Yamada1991,Chung2012}.
However, theoretical investigations~\cite{Huang2014,Chabot2003,
Yu2013, DaSilva2015} consistently find
the \Ba \ phase to be unstable against spontaneous rotation of the 
SnI$_6$ octahedra, so on energetic grounds
this phase
should not exist at all.
The answer to this puzzle must partly lie in the fact that the \Ba \ phase
is stable only at high temperature~\cite{Chung2012}, where both energetic
and entropic contributions determine the free energy $F$.
Unfortunately the most widely-used approach of calculating $F$ from first principles,
the quasiharmonic approximation~\cite{Baroni2001}, cannot be 
straightforwardly applied~\cite{DaSilva2015} due to the presence of the unstable (imaginary) 
phonon modes (Fig.~\ref{fig.harm_comp}).

In this Letter, we demonstrate the critical role played by anharmonicity
in stabilizing the experimentally-observed cubic and tetragonal
(\Ba \ and \Bb) phases of \CSI.
We perform our \emph{ab initio} investigation 
using a stochastic implementation
of self-consistent phonon theory~\cite{Hooton1955,Gillis1968,Brown2013,Errea2014}.
We show that the SnI$_6$ octahedra are stabilized against tilts and rotations
by interacting with the renormalized vibrations of the Cs ions.
Unexpectedly our calculations also reveal a temperature-induced
opening of the band gap, with a magnitude of 0.24 and 0.11~eV at 500~K and
300~K, respectively.
The significant size of these corrections (36~and 11\% of the uncorrected gaps)
places temperature effects
at a similar level of importance as spin-orbit
coupling for determining the band gap in these materials~\cite{Umari2014}, 
yet usually they are not included in \emph{ab initio} studies.
We further find that the gap-opening is not consistent with
a harmonic theory of band gap renormalization~\cite{Allen1976}, but
can be understood in terms of an increase in
average length of the Sn--I bonds.

All total energy and force calculations in this work were performed within a 
generalized-gradient approximation to density-functional theory (the PBEsol functional~\cite{Perdew2008}),
expanding the wavefunctions in a plane-wave basis 
set~\footnote{
We used plane-wave cutoffs of 1000~and~600~eV for the phonon and band gap calculations.
We used supercells and ($\Gamma$-centred)
$k$-point grids of 
2$\times$2$\times$2/4$\times$4$\times$4 (\Ba),
1$\times$1$\times$2/6$\times$6$\times$4 (\Bb), and
1$\times$1$\times$1/6$\times$4$\times$6 (\Bg).} 
and
treating the interactions between electrons and ion cores
within the projector-augmented wave formalism~\cite{Blochl1994} as implemented in the \texttt{GPAW} 
code~\cite{Enkovaara2010}.

\begin{figure}
\includegraphics{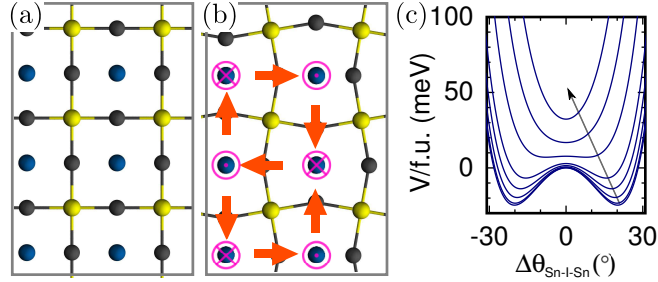}
\caption{(color online)
(a)  \Ba \ phase of \CSI, with blue/yellow/grey atoms = Cs/Sn/I.
(b)  \Ba \ structure distorted along modes at $M$
point corresponding to SnI$_6$ octahedral rotation (orange arrows)
and Cs vibration (pink arrows in/out of page).
The respective amplitudes along each mode are $x_\mathrm{rot}$ and $x_\mathrm{Cs}$.
(c) Energy $V$ vs deviation in Sn--I--Sn bond angle $\theta$
for constant values of $x_\mathrm{Cs}$ of 
(in direction of arrow) 0.0, 0.25, 0.5, 0.75, 1.0, 1.5, 2.0, 2.5, in units of
$\sqrt{\<x_\mathrm{Cs}^2\>}_{T=500K}$.
\label{fig.ball_stick}
}
\end{figure}
In Fig.~\ref{fig.harm_comp} we show
the phonon bandstructure obtained for the \Ba \ phase
calculated in the harmonic approximation using the finite displacement method~\cite{Ackland1997,Maradudin1968}.
Instabilities corresponding to tilts and rotations of the SnI$_6$ 
octahedra are found
at the $M$ and $R$ points of the Brillouin zone~\cite{Huang2014}.
One of the triply-degenerate soft
modes at $M$ is shown in Fig.~\ref{fig.ball_stick},
together with a frozen-phonon calculation of the potential energy surface (PES)
with respect to a distortion $x_\mathrm{rot}$ along this mode.
The cubic structure ($x_\mathrm{rot}=0$) is metastable,
and the system can lower its potential energy through an octahedral rotation to
a new structure with tetragonal symmetry.
These soft modes have been observed under a number of different 
computational setups and theoretical approximations~\cite{Huang2014,Chabot2003,Yu2013,DaSilva2015}, 
and cannot be stabilized for example by the application of a strain.
In fact increasing the lattice constant above its 0~K equilibrium value yields further soft modes, 
identified as ferroelectric instabilities in Ref.~\citenum{DaSilva2015}.

The fact that experiments observe only the cubic phase 
at temperatures above 440~K~\cite{Scaife1974,Yamada1991,Chung2012}
indicates that this phase corresponds to a minimum
of the free energy $F$.
In the quasiharmonic approximation~\cite{Baroni2001}, $F$ is replaced with $\F(\wall,T)$, the
free energy of an ensemble of oscillators of temperature $T$ with frequencies
$\wall = \{\w_1,\w_2,...\w_\nu\}$:
\begin{equation}
\F(\wall,T) = V_0 + \sum_\nu \left[ \frac{\hbar\w_\nu}{2} - k_B T \ln [1 + n_B(\w_\nu,T)] \right].
\label{eq.F_SHO}
\end{equation}
$V_0$ is the energy of the ions in their equilibrium positions,
$\hbar$ and $k_B$ the Planck and Boltzmann constants, and $n_B$ the Bose-Einstein distribution
function.

A quasiharmonic treatment of the \Ba \ phase would replace $\wall$ with the phonon
eigenfrequencies shown in Fig.~\ref{fig.harm_comp},
but there are two difficulties:
First, equation~\ref{eq.F_SHO} is defined only for real phonon frequencies,
so the contribution to $F$ from the soft modes cannot be included.
Second, from the harmonic phonon frequencies and eigenvectors of Fig.~1
we calculate that Cs atoms would undergo typical oscillations with 
a root  mean-square displacement of 0.8~\AA \ at 500~K, corresponding 
to over 18\% of the distance to their iodine neighbors at equilibrium~\cite{suppinfo}.
Such large displacements are unlikely to be well-described within the harmonic approximation.

Determining $F$ for the \Ba \ phase therefore requires moving
beyond the (quasi)harmonic approximation.
Different approaches to this problem have been developed, including
methods based on parametrizations and perturbative expansions of the 
PES~\cite{Zhong19942,Thomas2013,Monserrat2013},
molecular dynamics~\cite{Hellman2011,Zhang2014}, and self-consistent
phonons~\cite{Brown2013,Errea2014,Hooton1955,Gillis1968,Souvatzis2008}.
Here we follow the self-consistent phonon approach and calculate a fictitious free 
energy $\Ff(\wall,T)$ as
\begin{equation}
\Ff(\wall,T) = \F(\wall, T) + \<V\>_T - \<\V\>_T.
\label{eq.F_SCw}
\end{equation}
Here $\V$ is a harmonic approximation to the true
PES $V$, and $\<A\>_T$ is a thermal average of a quantity $A$ with respect
to the fictitious harmonic system, whose exact value is obtained via Mehler's 
formula~\cite{Watson1933,Patrick20142} as
\begin{equation}
\<A\>_T = \prod_\nu \frac{1}{\sqrt{2\pi \<x_\nu^2\>_T}} \int d x_\nu \
e^{-x_\nu^2/2\<x_\nu^2\>_T}
A(\vec{x}).
\label{eq.therm_av}
\end{equation}
$\vec{x}$ gives the amplitudes along each phonon mode $\nu$,
with the mean-square amplitude at temperature $T$ given as
$\<x_\nu^2\>_T$.
$\Ff$ is the free energy of the real system evaluated on the thermal equilibrium
state of the fictitious system, and is a rigorous upper bound to the true free energy $F$~\cite{Errea2014}.
The self-consistent set of frequencies $\wall$ are chosen as those
which minimize $\Ff$.

A fully self-consistent phonon theory (e.g.~Refs.~\citenum{Brown2013,Errea2014})
also minimizes $\Ff$ with respect to phonon eigenvectors and equilibrium ionic positions,
but in the current study we keep these quantities fixed at their harmonic values.
The reasons for performing this simplification are (i) for the high-symmetry \Ba \ phase,
many of the phonon eigenvectors (including the soft modes)
are fixed by the crystal symmetry, and 
(ii) the large unit cells and low symmetry of the \Bb~and~\Bg \ phases
render a full minimization of $\Ff$ impractical~\footnote{
As an example, a 1$\times$1$\times$1 sampling for the \Bg \ phase
would require minimizing $\Ff$ with respect to 861 basis functions, 
compared to the 25 realized in
Ref.~\citenum{Errea2014}.},
even after 
performing the symmetrization techniques of Ref.~\citenum{Errea2014}.
Then, as has been done previously for calculating free energies~\cite{Brown2013,Errea2014}, 
absorption spectra~\cite{Patrick2013}
and magnetic spectroscopies~\cite{Rossano2005}
we evaluate the thermal averages of equation~\ref{eq.therm_av} 
stochastically from an ensemble of configurations with ionic displacements
distributed according to $\prod_\nu \exp[-x_\nu^2/(2\<x_\nu^2\>_T)]$.
We label the current scheme \SCw.

Figure~\ref{fig.harm_comp} shows the \SCw-calculated 
phonon bandstructure obtained at 500~K.
There are three points to note.
First, the soft modes at the $M$ and $R$ points are stabilized to positive
energies of 2.3~meV.
Second, the vibrational energies of the Cs atoms appearing at 1--3~meV
in the harmonic approximation~\cite{DaSilva2015} are renormalized by 
more than a factor of two in \SCw.
As a result, ferroelectric instabilities involving Cs atoms 
that appear at a strained
lattice vanish at high temperatures~\cite{suppinfo}.
Finally, the lattice constant which minimizes $\Ff$ is calculated to be
6.21~\AA, which compared to experiment (6.206~\AA~\cite{Chung2012}) is
a significant improvement over the values of 6.131~\AA \ found
by minimizing the total energy, and 6.160~\AA \ obtained at 500~K
from a quasiharmonic analysis
ignoring the soft modes~\cite{DaSilva2015}.

The significant renormalization of the Cs vibrations
points to the mechanism by which
the soft modes are stabilized in \SCw.
The \SCw \ potential calculated for octahedral rotations is far
steeper than that expected from a one-dimensional analysis of
a quartic potential, which yields a parabola
wide enough for the system to sample the two minima~\cite{suppinfo}.
Instead one must consider phonon-phonon interactions  
between the octahedral rotations and the 
vibrations of the Cs atoms.
In Fig.~\ref{fig.ball_stick}(c) we show
the PES obtained by simultaneously displacing the Cs atoms along
an $M$-point phonon whilst rotating the SnI$_6$ octahedra.
Harmonically for each Cs mode amplitude $x_\mathrm{Cs}$ one
would expect an identical PES, offset by an energy $1/2 M_P \w_\mathrm{Cs}^2 x_\mathrm{Cs}^2$
($M_P$ is the proton mass).
Instead, the PES changes shape, showing that terms  like $ x_\mathrm{Cs}^2 x_\mathrm{rot}^2$
stabilize the cubic structure.
We stress that the analysis of Fig.~\ref{fig.ball_stick}(c) only
couples two phonon modes, whilst \SCw \ includes all couplings.

Crucially, the value of $\Ff$ calculated for the cubic phase
is 20~meV per formula unit lower than the tetragonally-distorted
phase at 500~K, showing that at high temperature
it is more beneficial to the free energy to have the Cs atoms
vibrating in a large volume than it is to reduce $V_0$ by rotating
the SnI$_6$ octahedra.
Our calculations corroborate the experimental
interpretation of  Cs atoms ``rattling''
within the perovskite cages~\cite{Chung2012}.

Given the interest in the optoelectronic properties of \CSI,
it is desirable to quantify the effects of phonons on the 
electronic band gap $E_g$.
There is increasing evidence that semilocal
exchange-correlation functionals find a weaker electron-phonon coupling
strength compared to more sophisticated theories of electronic
excitations, e.g.\ the $GW$ approximation~\cite{Antonius2014,Faber2015}.
For this reason we perform electronic structure calculations using the 
derivative discontinuity-corrected
GLLB-SC functional of Ref.~\citenum{Kuisma2010},
which has been found to improve the PBEsol description of the band 
gap for a range of materials~\cite{Li2015,Castelli2012}.
We calculate a gap deformation potential of 7.20~eV with GLLB-SC,
close to the value of 7.35~eV found from the quasiparticle self-consistent
$GW$ (QSGW) calculations of Ref.~\citenum{Huang2013} and steeper than the
values of 4.73 found with PBEsol or 4.65~eV  from the local-density approximation~\cite{Huang2013}.
The derivative discontinuity is responsible for this difference~\cite{suppinfo}.

\begin{figure}
\includegraphics{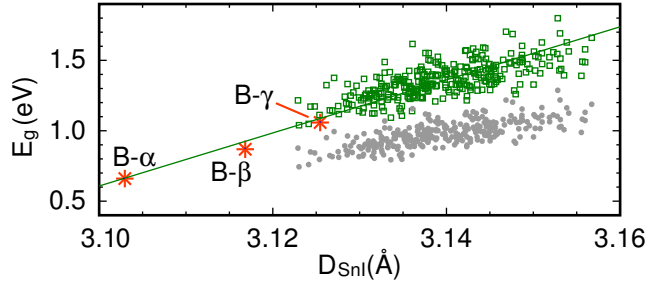}
\caption{(color online)
GLLB-SC band gaps calculated with/without (green squares/grey circles)
derivative discontinuity contribution for an ensemble of 300 configurations
vs average Sn--I bond length $D_{\mathrm{SnI}}$.
A linear fit to the data is shown.
$E_g$ is taken as the difference between the highest occupied state
and the average of the three lowest unoccupied states 
at $R$~\cite{suppinfo}.
We also show the gaps of the unperturbed \Ba, \Bb \ and \Bg \ structures (orange stars).
Note these calculations were performed in a 2$\times$2$\times$2 supercell and
subject to the finite size effects discussed in the text.
\label{fig.gap}
}
\end{figure}
Expanding the lattice constant from 6.131~\AA \ (harmonic, $T$=0~K)
to 6.21~\AA \ (\SCw, $T$=500~K) already accounts for an increase
of the gap $E_g$ from 0.40~eV to 0.66~eV.
However in addition there is a constant-volume
renormalization of the gap due to phonons~\cite{Cardona2005}, which in the adiabatic approximation
of Ref.~\citenum{Allen1976} is calculated as $\Delta E_g = \<E_g\>_T - E_g^0$,
where $E_g^0$ is the gap calculated with the ions in their equilibrium
positions~\cite{Patrick20142}.
We use equation~\ref{eq.therm_av} to evaluate $\<E_g\>_T$ from the \SCw \ frequencies
at the experimental volume at 500~K.
The band gaps calculated for 300 configurations
is shown in Fig.~\ref{fig.gap}.
The calculated $\Delta E_g$ is remarkable for being both large
and positive, i.e.\ the electron-phonon interaction increases the gap.
Although the latter behavior has been observed experimentally for materials like copper
halides~\cite{Cardona2005,Serrano2002}, \emph{ab initio} calculations of 
electron-phonon renormalization have so far focused on semiconductors like diamond and 
Si where the gap is reduced
by temperature~\cite{Giustino2010,Cannuccia2011,Antonius2014,Ponce2015}.

We have also studied the technologically-relevant low 
temperature \Bb~and~\Bg \ phases at 380 and 300~K,
respectively.
Owing to the close agreement of the \SCw \ \Ba \ 
lattice constant with experiment, we used the 
experimental lattice constants 
reported in Ref.~\citenum{Chung2012} for the other phases.
We show the \SCw \ bandstructures in the supplemental information~\cite{suppinfo}.
For the \Bb \ phase, the \SCw \ calculations remove the unstable
modes and renormalize the frequencies of the 
Cs modes, whilst
for the \Bg \ phase some small
changes in phonon frequencies occur across the spectrum~\cite{suppinfo}.
The calculated corrections to the band gap are again large,
yielding $+0.70$~eV (\Ba \ phase, 500~K),
$+0.45$~eV (\Bb \ phase, 380~K) and $+0.31$~eV (\Bg \ phase, 300~K).
However as discussed below these values are likely to be overestimates due 
to finite size effects in our supercell calculations.

\begin{figure}
\includegraphics{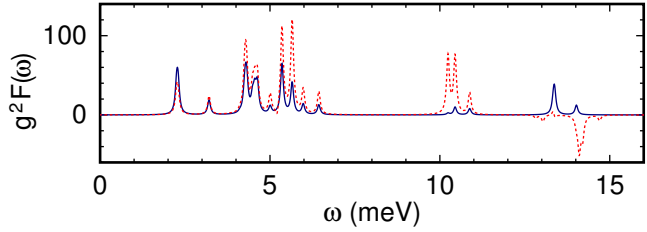}
\caption{
(color online)
Gap spectral functions $g^2F(\w)$ calculated using
$\partial E_g/\partial n_\nu = l_\nu^2 \partial^2 E_g / \partial x_\nu^2$ (red dotted line)
and equation~\ref{eq.resum} (blue line).
The expected gap renormalization at temperature $T$ is obtained 
as $ \int d\w g^2 F(\w) [n_B(\w,T) + 1/2]$.
\label{fig.specfunc}
}
\end{figure}
To further investigate this band gap renormalization we first consider the harmonic theory of Ref.~\citenum{Allen1976}, where 
\begin{equation}
\Delta E_g \approx\Delta E_g^{(2)} = \sum_\nu \frac{\partial E_g}{\partial n_\nu} \left[ n_B(\w_\nu,T) + \frac{1}{2} \right].
\label{eq.dedn}
\end{equation}
Usually the coupling coefficient $\partial E_g/\partial n_\nu$ is defined as 
$l_\nu^2 \partial^2 E_g / \partial x_\nu^2$ 
with $l_\nu$ being the characteristic length of the normal mode~\cite{suppinfo},
but with this definition equation~\ref{eq.dedn}
gives a too large gap renormalization of 0.95~eV at 500~K,
demonstrating the failure of a harmonic expansion of $E_g$ with respect to $x_\nu$.
However Fig.~\ref{fig.gap} reveals a correlation between the calculated
gap and $D_{\mathrm{SnI}}$, the average Sn--I bond length,
which accounts for the gap increase both from the electron-phonon interaction 
and between the unperturbed $\alpha$, $\beta$ and $\gamma$ phases (orange stars).
Following Ref.~\citenum{Huang2013} we attribute this sensitivity 
to a weakened Sn-$s$/I-$p$ antibonding interaction as the bond length increases,
narrowing the valence band and widening the band gap.
This correlation motivates a resummation and new coupling constant definition:
\begin{equation}
\frac{\partial E_g}{\partial n_\nu} = l_\nu^2 \frac{dE_g}{d D_{\mathrm{SnI}}} \frac{\partial^2 D_{\mathrm{SnI}}}{\partial x_\nu^2}
\label{eq.resum}
\end{equation}
where $dE_g/d D_{\mathrm{SnI}}$ is the gradient of the straight line in Fig.~\ref{fig.gap}.
In Fig.~\ref{fig.specfunc} we plot the spectral functions~\cite{Capaz2005}
$ g^2F(\w) = \sum_\nu \partial E_g/\partial n_\nu \delta(\w-\w_\nu)$ for the two definitions of 
$ \partial E_g/\partial n_\nu$, showing that
(a) the harmonic expansion of $E_g$ predicts much larger contributions from polar modes at 6~meV,
and (b) both expansions yield an important contribution to the gap renormalization from the octahedral rotations 
at 2.3~meV, which can only be described with an anharmonic treatment of the ground state.

Equation~\ref{eq.resum} yields a gap
renormalization of 0.70~eV for the \Ba \ phase at 500~K, exactly reproducing the ensemble
average of Fig.~\ref{fig.gap}.
Noting that the \Bb \ and \Bg \ phases display a similar correlation of $E_g$ with
$D_{\mathrm{SnI}}$~\cite{suppinfo}, we combined
$dE_g/dD_{\mathrm{SnI}}$ from Fig.~\ref{fig.gap} with 
$\partial^2 D_{\mathrm{SnI}}/\partial x_\nu^2$ obtained
from the phonon eigenvectors of these phases and found
renormalizations of 0.47 and 0.32~eV, also remarkably consistent with
the full ensemble averages.

The surprisingly large band gap corrections 
raise two questions, first whether the adiabatic 
interpretation of $\<E_g\>_T$ as the electron-phonon-corrected
band gap~\cite{Allen1976} is sufficient to describe the photophysics of this polar 
material~\cite{Ponce2015,polarnote},
and second
whether the supercells used to calculate the gap renormalization
have introduced finite size effects (e.g.\ through an oversampling of the soft modes).
Current methods of treating non-adiabaticity 
have not yet been extended to systems dominated by
anharmonic couplings between different phonon modes~\cite{Antonius2015, Ponce2015},
but we studied the finite size effect
by Fourier-interpolating the \SCw \ dynamical matrix to
progressively larger $N$$\times$$N$$\times$$N$ supercells of the \Ba \ phase
and repeating the sampling of the band gap, utilizing
the localized-orbital basis sets implemented in \texttt{GPAW}~\cite{Larsen2009,suppinfo}.
We indeed observe slow convergence with supercell size, with an empirical 1/$N$ scaling.
Extrapolating this behavior
leads to a significant reduction of $dE_g/dD_{\mathrm{SnI}}$  by 62\%,
thus giving revised estimates of the gap renormalization from equations~\ref{eq.dedn}
and~\ref{eq.resum} of +0.24, +0.16 and +0.11~eV for \CSI \ 
at 500, 380 and 300~K, or corrected GLLB-SC gaps of 0.90, 1.04 and 1.17~eV.
Future work is required to study the nature and origin of this slow size convergence.

Connecting our work to experimental studies, 
we note that Ref.~\citenum{Yu2011} 
found the peak photoluminescence (PL) to increase in energy by 0.09~eV
from 9 to 300~K.
Although this data appears to agree with our calculated shift of +0.11~eV,
we note that (a) the latter value does not include thermal expansion effects,
and (b) it is unclear whether the PL corresponds to band--band 
transitions or defects~\cite{Chung2012,Xu2014}.
At higher temperatures, our calculations indicate that the band gap will reduce  
e.g.\ by 0.14~eV between 380 and 500~K.
The measurement of the absorption spectrum of \CSI \ 
over the 0--500~K temperature range would be highly useful to further explore these effects.

Finally, we note that whilst anharmonicity has been demonstrated to play a crucial
role for materials at very high temperatures or pressures~\cite{Errea2014,Souvatzis2008}, 
the conditions simulated here are relevant to the expected operating 
conditions for solar cells~\cite{Jones2001}.
It is notable that the 0.24~eV shift obtained for the cubic phase is of similar magnitude
to the spin-orbit correction~\cite{Huang2013}, with opposite sign.
Our work thus illustrates the importance of anharmonic temperature effects
to the realistic modeling of the X(Sn,Pb)Y$_3$ perovskites.

\begin{acknowledgments}
We thank I.\ Errea, I.E.\ Castelli and J.M. Garc\'ia-Lastra for useful
discussions, and acknowledge support from the Danish Council for 
Independent Research's Sapere Aude Program, Grant No. 11-1051390.
\end{acknowledgments}

%
\end{document}